\documentclass[aps,prd,nofootinbib]{revtex4}
\usepackage{graphicx}
\usepackage{amsfonts}
\usepackage{amsmath}
\usepackage{amssymb}

\begin{document}

\title{Damour--Solodukhin Wormhole as a Black Hole Mimicker: The Role of Observers' Location}
\author{K.K. Nandi}
\email{kamalnandi1952@rediffmail.com}
\affiliation{High Energy Cosmic Ray Research Center, University of North Bengal, Siliguri 734013, WB, India}
\affiliation{Zel'dovich International Center for Astrophysics, Bashkir State Pedagogical University, 3A, October Revolution Street, Ufa 450008, RB, Russia}
\affiliation{Department of Physics \& Astronomy, Bashkir State University, 47A, Lenin Street, Sterlitamak 453103, RB, Russia}
\author{R.Kh. Karimov}
\email{karimov_ramis_92@mail.ru}
\affiliation{Zel'dovich International Center for Astrophysics, Bashkir State Pedagogical University, 3A, October Revolution Street, Ufa 450008, RB, Russia}
\author{R.N. Izmailov}
\email{izmailov.ramil@gmail.com}
\affiliation{Zel'dovich International Center for Astrophysics, Bashkir State Pedagogical University, 3A, October Revolution Street, Ufa 450008, RB, Russia}
\author{A.A. Potapov}
\email{a.a.potapov@strbsu.ru}
\affiliation{Department of Physics \& Astronomy, Bashkir State University, 47A, Lenin Street, Sterlitamak 453103, RB, Russia}

\date{09 October 2022}

\begin{abstract}
It has been recently argued that in semi-classical gravity, a minimal 2-sphere is not a horizon but a tiny throat of a wormhole, such as the Damour--Solodukhin wormhole (DSWH), with a free parameter $\lambda \neq 0$ separating it from a Schwarxzschild black hole (BH) ($\lambda =0$). As shown by DS, their horizonless WH can mimic many properties of a black hole (BH). Assuming that observing a BH mimicker is equivalent to observing a BH itself, we ask the question as to which identity of the object, a WH or a BH, an observer is likely to observe in a single experiment. To answer this, we introduce Tangherlini's new concept of indeterminacy in the gravitational field by portraying the field as a refractive medium. We then postulate that \textit{the identity of the observed object will depend on the probabilistic outcome of photon motion probing the object}.  The probabilities will be described by Fresnel reflection ($R$) and transmission ($T$) coefficients derived by Tangherlini on the basis of a non-quantum statistical indeterminacy of photon motion in ordinary optical media. By adapting this approach to a gravitational "effective optical medium," we obtain two intriguing results: (i) The Fresnel coefficients at the DSWH throat are independent of mass $M$ but dependent solely on the parameter $\lambda \neq 0$. (ii) Depending on the location of the observer, what is a DSWH to one observer may appear as a BH to another observer for the same value of $\lambda \neq 0$.
\end{abstract}

%\pacs{}
\maketitle

%%%%%%%%%%%%%%%%%%  DATE  %%%%%%%%%%%%%%%%%%%

%%%%%%%%%%%%%%%%%%%%%%%%%%%%%%%%%%%%%%%%%
\section{Introduction}
\label{sec:intro}
%%%%%%%%%%%%%%%%%%%%%%%%%%%%%%%%%%%%%%%%%
Recently, Barthiere, Sarkar and Solodukhin \cite{Barthiere:2018} proved a remarkable result, viz., in semi-classical gravity with conformal fields propagating on a classical background, a static, spherically symmetric metric with a minimal 2-sphere is not a horizon but a tiny throat of a wormhole. A prototype of such an object is the Damour--Solodukhin wormhole (DSWH) \cite{Damour:2007} with a parameter $\lambda$ which separates the WH ($\lambda \neq 0$) from the Schwarzschild black hole (BH)\ ($\lambda =0$). Black holes and wormholes possess different topologies, and while the formation of BHs is understood to be a possible end product of material collapse under appropriate conditions or even as a result of WH collapse \cite{Gonzalez:2009, Bronnikov:2011, Bronnikov:2012, Bronnikov:2004, Bronnikov:2007}, the mechanism for the formation of classical wormholes is still not well understood. While BHs have been definitively observed to exist in nature (e.g., the supermassive SgrA* in the Milky Way center), WHs are still largely speculative but have not yet been ruled out by experiments. They are valid solutions of Einstein's equations just as BHs are, and thus legitimately occupy an important place in physics, such as in \cite{Barthiere:2018, Damour:2007}.

Apart from Einstein's theory, there is already a vast literature of extended and alternative theories of gravity admitting BHs and WHs. It is impossible to exhaustively list them all here---however, some random examples are WHs in the $f(R,T)$ theories of modified gravity \cite{Sahoo:2021}, in $f(R)$ gravity \cite{Bhattacharya:2017}, in alternative theories \cite{Sahoo:2018}, in Einstein--Born--Infeld theory \cite{Richarte:2009a} and in string theories \cite{Nandi:1998a}. The most well known alternative theory, the Brans--Dicke theory \cite{Nandi:1997, Agnese:1995}, and its conformal re-incarnation, the Einstein minimally coupled theory \cite{Nandi:1998b}, both admit traversable WHs and BHs. Reasonably detailed discussion of WHs in exotic theories is available also in \cite{Moffat:2020}. Examples of BHs in modified theories can be found \mbox{in \cite{Lessa:2020}}.

The DSWH metric differs from the Schwarzschild BH metric by a tiny parameter, $\lambda \neq 0$, so that instead of the BH horizon, the DSWH has a throat. A throat is the radius of the minimal 2-surface so that the WH does not have a center, unlike BH. Despite topological differences between these two objects, the horizonless DSWH surprisingly mimics many features of a BH (endowed with an event horizon), as long as the WH parameter $\lambda $ is so tiny that the corresponding travel time is too large, i.e., in the order of $\Delta t\sim 2GM\ln \left( \frac{1}{\lambda ^{2}}\right) $ \cite{Damour:2007}. One knows that (classically) light cannot emerge from a BH horizon, or said differently, light from the horizon takes infinite time to reach an asymptotic observer. This behavior of light from the horizon could be effectively similar to the behavior of a DSWH throat in that the signals emerging from it cannot be received by an asymptotic observer if the travel time $\Delta t$ to reach the reception point is too large because of a too-small $\lambda \neq 0$. This is an example of how a WH throat can mimic the behavior of a BH horizon.

As discussed by Damour and Solodukhin \cite{Damour:2007}, the mimicking features include the accretion disc, no-hair properties, quasi-normal-mode ringing and even the dissipative properties of black hole horizons. Study of Page--Thorne thin-accretion-disk profiles showed that WHs, BHs and naked singularities become almost indistinguishable in the far-field observations, though they may greatly differ in the near-source limit \cite{Karimov:2020}. It was also shown in \cite{Karimov:2019} that even the rotating DSWH are experimentally indistinguishable from Kerr BHs in the far field. As shown in \cite{Nandi:2018}, the strong-field lensing properties of DSWH mimic those of Schwarzschild BH within the bound $\lambda \leq 10^{-3}$. (Note that this is just an upper bound, which does not preclude any finite non-zero value of $\lambda$ arbitrarily smaller than that). Of course, there are many theoretical works distinguishing BHs and WHs, and of particular interest are the works distinguishing Kerr-like Damour WHs and Kerr BHs using different methods \cite{Kasuya:2021, Kiczek:2022, Amir:2019}. In these studies, one considers two metrics of WH and BH spacetimes to compare the observables. Here we are concerned with only the single metric of DSWH and its BH mimicking, even when  $\lambda \neq 0$. We assume that observing a BH mimicker is equivalent to observing the BH itself.

The purpose of the present paper is to address the question as to which identity of the gravitating object, a DSWH or a mimicked BH,  an observer is likely to observe in a single experiment. It is exactly here that we introduce a novel concept of indeterminacy, due to Tangherlini \cite{Tangherlini:1975}, in the identity of a probed gravitating object that has not been studied heretofore except in \cite{Nandi:2016}, to our knowledge. Tangherlini's introduced a non-quantum indeterminacy in the ordinary optical medium to derive Fresnel reflection ($R$) and transmission ($T$) coefficients for photons. We adapt this idea to the "effective optical medium" equivalent to DSWH spacetime for the probabilistic outcome of probe by photons---i.e., we calculate the Fresnel coefficients ($R,T$) for the DSWH, which yield the probability of its behaving as a WH \textit{or} a BH mimicker. \textit{We take the values} $R=1, T=0$ \textit{to represent a certain outcome to be a Schwarzschild BH, for which }$n(r)\rightarrow \infty$ \textit{at the horizon, which at once yields} $R=1$, $T=0$. Any deviation from these values preserving $R+T=1$ would represent the outcome to be a WH or BH with corresponding probabilities, and the other extreme, $R=0,T=1$, would represent the certain outcome of a WH.

Photons sent into an "effective optical medium" by an asymptotic observer will sense a refractive index $n(r)$, and photons sent by near-throat observers will sense another index, $\widetilde{n}(r)$\footnote{%
Yet another different index $N(r)$ is sensed by a massive particle in motion \cite{Nandi:2016, Evans:2001, Evans:1996a, Evans:1996b, Alsing:2001, Alsing:1998}. In the present paper, we deal with only photon motion while taking into account that, in general relativity, observations depend on the location of the observer. The photon's motion in the effective medium also gives rise to the interesting possibility of the gravitational analogue of the Fizeau effect in the "effective optical medium" \cite{Nandi:2003}.} \cite{Nandi:1995}. Stated differently, asymptotic observers will measure $n(r)$, and near-throat observers will measure $\widetilde{n}(r)$. We emphasize that the present application of Tangherlini's formulation to the gravitational "effective optical medium" is no blind application, but is solidly rooted in the famous Pound--Rebka experiment in a gravity field. The validity of this assertion can be seen from the following argument: Tangherlini's formulation starts with the photon momentum increase according to the law $p^{\prime}=np$, when it travels into a denser medium characterized by a refractive index $n>1$. While this increase is already an experimentally confirmed fact in an ordinary optical medium \cite{Poynting:1905, Jones:1951, Jones:1954, Ashkin:1973}, the Pound--Rebka experiment can be re-interpreted as yet another parallel experiment in the "effective optical medium"; and index $n(r)$ remarkably confirms exactly the same momentum increase law, as shown in detail in \cite{Nandi:2016}.

The paper is organized as follows: In Section \ref{sec:T_interpr}, a brief review of Tangherlini's formulation is presented, together with the expressions for Fresnel coefficients ($R$ and $T$) in the "effective refractive medium" as perceived by two different observers, both probing the throat by sending photons. Section \ref{sec:app_to_DSWH} applies them in the "effective optical medium" corresponding to DSWH. Section \ref{sec:concl} concludes the paper.

%%%%%%%%%%%%%%%%%%%%%%%%%%%%%%%%%%%%%%%%%
\section{Tangherlini's Formulation of $R$ and $T$}
\label{sec:T_interpr}
%%%%%%%%%%%%%%%%%%%%%%%%%%%%%%%%%%%%%%%%%
Tangherlini's \cite{Tangherlini:1975} method depends in part on the increase in momentum of the photon, when it travels from free space into a plane, semi-infinite, homogeneous, isotropic, non-absorbing ordinary optical medium, and in part on the assumption of a probabilistic condition at the surface where the photon impinges. If the photon is transmitted, the magnitude of its momentum $p^{\prime }$ in the medium is related to the magnitude of its momentum $p$ in free space by the equation
\begin{equation}
p^{\prime }=np,
\end{equation}%
where $n$ ($\geq 1$) is the index of refraction of the medium measured by an asymptotic observer. This equation, which holds independently of the angle of incidence, is an experimental fact in an ordinary medium \cite{Poynting:1905, Jones:1951, Jones:1954, Ashkin:1973}. Note that the coordinate de Broglie relation
\begin{equation}
p^{\prime}\lambda^{\prime} = p\lambda = \text{constant}
\end{equation}%
(without Planck's constant), together with Equation (1), leads to%
\begin{equation}
\lambda^{\prime }n = \lambda = \text{constant},
\end{equation}%
where $\lambda^{\prime}$ and $\lambda$ are the wavelengths in the optical medium and in free space, respectively. Since Planck's constant is absent in Equation (2), Tangherlini called it a "non-classical, pre-quantal" description of light motion in the ordinary optical medium.

From a dynamical standpoint, the treatment of a photon may be based in the ray approximation on the following Hamiltonian, for negligible dispersion \cite{Tangherlini:1975}:%
\begin{equation}
H^{\prime }=\frac{c_{0}}{n(r)}p^{\prime },
\end{equation}%
where $c_{0}$ is the vacuum speed of light and $n=n(r)$ is a slowly varying refractive index for a real medium so oriented that the boundary is in the transverse plane. Note that gravity does not as yet play any role---all quantities $p^{\prime}$, $p$, $\lambda^{\prime}$, $\lambda$, and $n$ are coordinate quantities in a Euclidean space that are measured by scales and clocks unaffected by gravity. \textit{Remarkably, all of the above relations can be exactly derived for photon or particle motion in a static spherically symmetric gravitational field when portrayed as an "effective optical medium" with gradient index} $n(r)$ (see, for details, \cite{Nandi:2016}). An application of the Fermat principle
\begin{equation}
\delta \int_{\mathbf{x}_{1}}^{\mathbf{x}_{2}}n(r)\left\vert d\mathbf{r}\right\vert = 0,
\end{equation}%
where $\left\vert d\mathbf{r}\right\vert =\sqrt{dx^{2}+dy^{2}+dz^{2}}$ is the Euclidean distance, then reproduces all the exact geodesic equations of general relativity, as shown in \cite{Nandi:1995, Evans:2001, Evans:1996a, Evans:1996b}. Therefore, these formulas can be directly applied to DSWH spacetime as well, once we can find its associated $n(r)$. However, in the present work, we do not study geodesics but study the Fresnel coefficients resulting from the statistical indeterminacy of photon motion probing the identity of the gravitating object.

Tangherlini's \cite{Tangherlini:1975} ingenious idea of non-quantum indeterminacy is as follows. He replaced the classical deterministic wave propagation into optical media by introducing an \textit{indeterministic condition at the interface} achieved by a hypothetical arrangement in which individual photons are incident on widely separated replica media at random intervals of time. As a consequence, the physical flow of energy is now associated with some kind of a \textit{probability flow} associated with each photon. The Fresnel coefficients are then determined in terms of a statistical ensemble average over a large number of replica media having the same index $n$---with one photon for each replica. Due to the assumed independence of collisions at the medium interface, the coefficients are the same for one photon as for $N$ photons. The de Broglie relation $p^{\prime}\lambda ^{\prime} = p\lambda =$ constant (\textit{no Planck's constant} $\hslash$), together with well known reduction of wavelength $\lambda^{\prime} = \lambda/n$, lead to $p^{\prime} = np$, which was used as a starting point by Tangherlini to derive Fresnel coefficients $R$ and $T$. Remarkably, the same Fresnel coefficients can be deduced also from the standard quantum mechanical treatment of Schr\``{o}dinger's equation for a certain potential well, which reinforces the validity of the assumed probability flow in Tangherlini's scheme. Loosely speaking, "it is quantum mechanics without quantum mechanics." That is why Tangherlinini called his approach a "pre-quantal statistical formulation." The new idea in Tangherlini's treatment is that the photon motion in the medium is raised from the classical deterministic level to a so-called pre-quantal statistical flow of the energy \textit{sans} Planck's constant. For details, see the original paper \cite{Tangherlini:1975}.

To interpret Tangherlini's probabilistic idea in the gravitational "effective optical medium," we performed a parallel experiment in which a single photon was sent into the gravity field to probe the gravitating object. If all the photons are reflected back by the throat, the observer sending the pulses will determine $R$ $=1$, $T=0$ and identify the object with \textit{certainty} that it is a Schwarzschild BH, since for it, $n\rightarrow \infty$ at the horizon, yielding $R$ $=1$, $T=0$. On the other hand, any deviation from the values $R$ $=1$, $T=0$ (that is, some photons reflected, some transmitted) would mean that there is a \textit{non-zero probability} that the observer might end up identifying the probed object as a WH. Now suppose that there are two observers $A$ and $B$; their respective outcomes of the probe are $\left(R^{A},T^{A}\right)$ and $\left( R^{B},T^{B}\right)$, which differ from the values $(1,0)$ such that $R^{A}>R^{B}$. Then, the observer $A$ is more likely (but not certainly) to identify the object as a BH and the observer $B$ is more likely to observe it as a WH. Thus, when the two independent observers located far apart (asymptotic and near-throat) observe the same probe, it is likely that their probabilistic identification of the object will differ, one identifying it as a BH and the other identifying the same object as a WH. This should never be the case that for any $\lambda \neq 0$, however tiny; the ($R,T$) values at the throat measured by the two observers would  be exactly the same.

To formalize the above notions, let $I$ denote the incident particle flux, let $I_{R}$ be the flux of reflected particles, and let $I_{T}$ be the flux of transmitted particles such that
\begin{equation}
I_{R}+I_{T}=I.
\end{equation}%

In accordance with standard notation, the ratios $R= I_{R}/I$ and $T = I_{T}/I$, which we call Fresnel coefficients, define, respectively, the probabilities of reflection and transmission across any surface (here throat of the WH). They satisfy the conservation of probability condition%
\begin{equation}
R+T=1.
\end{equation}

Equating the average rate of energy delivered to each ensemble member in reflected and transmitted modes, Tangherlini \cite{Tangherlini:1975} derived the Fresnel coefficients as%
\begin{equation}
R = \frac{(n-1)^{2}}{(n+1)^{2}},\quad T = \frac{4n}{(n+1)^{2}}.
\end{equation}%
Note that $R$ and $T$ remain invariant under inversion of the index; $n\rightarrow 1/n$. These coefficients are the same as those that are obtained by solving the Schr\``{o}dinger equation for a specific potential \cite{Bohm:1951}. One could say that the quantum scenario with a specific potential has a classical analogue in Tangherlini's probabilistic thought experiment with many replica media.

Let us now consider a static spherically symmetric generic wormhole in isotropic coordinates
\begin{equation}
d\tau^{2} = \Omega^{2}(r)c_{0}^{2}dt^{2} - \Phi^{-2}(r)[dr^{2}+r^{2}\left(
d\theta^{2} + \sin^{2}{\theta }d\varphi ^{2}\right)],
\end{equation}%
where $c_{0}$ is the vacuum speed of light. This metric leads to a refractive index $n(r)$ for photon motion ($d\tau^{2} = 0$) defining the coordinate speed of light $c$ inside the "effective optical medium" measured by an \textit{asymptotic observer} (a.o.) as
\begin{eqnarray}
c &=&\left\vert \frac{d\mathbf{r}}{dt}\right\vert =\frac{c_{0}}{n(r)}, \\
n^{\text{a.o.}}(r) &=&\Omega ^{-1}\Phi ^{-1}.
\end{eqnarray}%

Note that $c$ is both the phase and group speed (see \cite{Nandi:2016}) in the "effective optical medium." Equation (11) defines the refractive index $n^{\text{a.o.}}$ observed by an asymptotic observer, since its length and time are unaffected by gravity, viz., those are $d\mathbf{r}$ and $dt$. The near-throat observers situated inside this medium measure, instead of the coordinate wavelength $\lambda^{\prime}$, a reduced proper wavelength following the metric (9) as
\begin{equation}
\widetilde{\lambda }=\lambda ^{\prime }\Phi ^{-1}
\end{equation}%
and so Equation (3) can be rewritten as%
\begin{eqnarray}
\widetilde{\lambda }(n\Phi ) &=&\widetilde{\lambda }\widetilde{n}=\lambda =%
\text{constant} \\
&\Rightarrow &\widetilde{n}^{\text{l.o.}}=n^{\text{a.o.}}\Phi =\Omega ^{-1}.
\end{eqnarray}%

Equation (13) is the proper version of Equation (3) yielding the refractive index $\widetilde{n}$ as measured by a \textit{local observer} (l.o.) inside the effective medium. Note that the \textit{proper version of the} de Broglie relation $\widetilde{p}\widetilde{\lambda} = p\lambda =$ constant (without Planck's constant) in the gravitational "effective optical medium" yields
\begin{equation}
\widetilde{p} = \widetilde{n}(r)p,
\end{equation}%
where $\widetilde{n}(r)>1$. Remarkably, it is precisely this momentum increase for an ingoing photon in an "effective optical medium" that has been supported by the Pound--Rebka experiment originally phrased as a gravitational frequency shift experiment \cite{Nandi:2016}.

The coefficients for the ingoing photon pulse at the throat, as observed by asymptotic observers (a.o.), are given, using Equation (8) with $n$:
\begin{eqnarray}
R_{\text{photon}}^{\text{a.o.}} &=&\frac{\left[ n(r_{\text{th}})-1\right]
^{2}}{\left[ n(r_{\text{th}})+1\right] ^{2}}, \\
T_{\text{photon}}^{\text{a.o.}} &=&\frac{4n(r_{\text{th}})}{\left[ n(r_{%
\text{th}})+1\right] ^{2}}.
\end{eqnarray}%

The coefficients for the ingoing photon pulse at the throat as observed by \textit{near-throat} local observers (l.o.) are given, using Equation (8) with $\widetilde{n}$ ($=\Omega ^{-1}$), by:
\begin{eqnarray}
\widetilde{R}_{\text{photon}}^{\text{l.o.}} &=&\frac{\left[ \widetilde{n}(r_{%
\text{th}})-1\right] ^{2}}{\left[ \widetilde{n}(r_{\text{th}})+1\right] ^{2}}%
, \\
\widetilde{T}_{\text{photon}}^{\text{l.o.}} &=&\frac{4\widetilde{n}(r_{\text{%
th}})}{\left[ \widetilde{n}(r_{\text{th}})+1\right] ^{2}}.
\end{eqnarray}%

The main idea in the next section is to apply the above coefficients in \mbox{Equations (16)--(19)}, which describe the probabilistic perception of the object---be it a WH or a BH---as perceived by the observers, depending on their locations.

%%%%%%%%%%%%%%%%%%%%%%%%%%%%%%%%%%%%%%%%%
\section{Application to a Damour--Solodukhin Wormhole}
\label{sec:app_to_DSWH}
%%%%%%%%%%%%%%%%%%%%%%%%%%%%%%%%%%%%%%%%%
A WH is a topological handle connecting two universes or two  distant regions of the same spacetime. The two distant regions are represented by two BHs with positive and negative masses, respectively. The positive mass attracts the observer, and the negative mass repels it. A traveler can pass from one region to the other by following the geodesic congruences connecting the two masses. Formal development of WH geometry can be found in the pioneering work of Morris and Thorne \cite{Morris:1988}, starting with a generic standard coordinate metric
\begin{equation*}
ds^{2}=e^{2\Phi (r^{\prime })}c_{0}^{2}dt^{2}-\left( 1-\frac{b(r^{\prime })}{%
r^{\prime }}\right) ^{-1}dr^{\prime 2}-r^{\prime 2}\left( d\theta ^{2}+\sin
^{2}{\theta }d\varphi ^{2}\right).
\end{equation*}%

The throat $r^{\prime }=r_{0}^{\prime }$ is defined by the root of the equation $b(r_{0}^{\prime })=r_{0}^{\prime }$.

A special case is the DSWH \cite{Damour:2007} metric given by
\begin{equation}
ds^{2}=\left( 1-\frac{2M}{r^{\prime }}+\lambda ^{2}\right) dt^{2}-\left( 1-%
\frac{2M}{r^{\prime }}\right) ^{-1}dr^{\prime 2}-r^{\prime 2}\left( d\theta
^{2}+\sin ^{2}{\theta }d\varphi ^{2}\right),
\end{equation}%
where $\lambda $ is the deviation parameter. To be traversable, the WH should not have a horizon, which is ensured by $\lambda \neq 0$. The throat of a wormhole is located at $r_{\text{th}}^{\prime }=2M$. At $\lambda =0$, the DSWH throat converts to Schwarzschild BH event horizon. For non-zero values of the parameter $\lambda $, the Einstein tensor of DSWH has a vanishing time--time component, while $G_{rr},G_{\theta \theta },G_{\phi \phi }\sim \lambda ^{2}$. This means, among other things, that matter with vanishing energy density would be required to sustain such a gravitational configuration.

Next, we will convert the solution to isotropic form using the transformation $r^{\prime}\rightarrow r$ as
\begin{equation}
r^{\prime} = r\left( 1+\frac{M}{2r}\right) ^{2}
\end{equation}%
so that the throat $r_{\text{th}}^{\prime }=2M$ converts to $r_{\text{th}} = M/2$. Then, the metric can be rewritten in the isotropic form
\begin{equation}
d\tau ^{2} = \left[ \left( \frac{1-\frac{M}{2r}}{1+\frac{M}{2r}}\right)^{2} + \lambda ^{2}\right] dt^{2}-\left( 1+\frac{M}{2r}\right)^{4}(dr^{2}+r^{2}d\Omega ^{2}),
\end{equation}%
where
\begin{eqnarray}
\Omega  &=&\left[ \left( \frac{1-\frac{M}{2r}}{1+\frac{M}{2r}}\right)^{2}+\lambda ^{2}\right] ^{1/2}, \\
\Phi  &=&\left( 1+\frac{M}{2r}\right) ^{-2}.
\end{eqnarray}%

We now explain the observers in the effective optical medium. Asymptotic observers (a.o.) are equipped with our instrumentations, and their observations coincide with those of ours. However, instrumentations and observations of observers inside the medium will be affected by gravitation, or equivalently, by the medium. For instance, the clock rate will change, and hence the wavelengths of ingoing photons near the throat will change according to $\lambda ^{\prime }=n\lambda =$ constant, as in Equation (3). Accordingly, the local observers (l.o.) near the throat will sense a refractive index $\overline{n}$ given by Equation (14). Any traveler passing through the WH and approaching the throat will perceive $\overline{n}$ and the corresponding Fresnel coefficients, Equations (18) and (19).

Therefore, the refractive index, as observed by asymptotic observers (a.o.), from \mbox{Equation (11)}, is
\begin{equation}
n^{\text{a.o.}}(r)=\left( 1+\frac{M}{2r}\right) ^{2}\left[ \left( \frac{1-%
\frac{M}{2r}}{1+\frac{M}{2r}}\right) ^{2}+\lambda ^{2}\right] ^{-1/2}
\end{equation}%
and similarly, the index as observed by near-throat local observers (l.o.), from \mbox{Equation (14)}, is
\begin{equation}
\tilde{n}^{\text{l.o.}}(r) = \left[ \left( \frac{1-\frac{M}{2r}}{1+\frac{M}{2r%
}}\right) ^{2}+\lambda ^{2}\right] ^{-1/2}.
\end{equation}%

At the throat $r_{\text{th}}=M/2$, the refractive indices have the values
\begin{equation}
n^{\text{a.o.}}(r_{\text{th}})=\frac{4}{\lambda },\quad \tilde{n}^{\text{l.o.%
}}(r_{\text{th}})=\frac{1}{\lambda}.
\end{equation}%

Using Equations (16)--(19), one has the Fresnel coefficients%
\begin{eqnarray}
R_{\text{photon}}^{\text{a.o.}} &=&\left( \frac{4-\lambda }{4+\lambda }%
\right) ^{2}, \\
T_{\text{photon}}^{\text{a.o.}} &=&\frac{16\lambda }{\left( 4+\lambda
\right) ^{2}},
\end{eqnarray}%
\begin{eqnarray}
\widetilde{R}_{\text{photon}}^{\text{l.o.}} &=&\left( \frac{1-\lambda }{%
1+\lambda }\right) ^{2}, \\
\widetilde{T}_{\text{photon}}^{\text{l.o.}} &=&\frac{4\lambda }{(1+\lambda
)^{2}}.
\end{eqnarray}

\begin{figure}
\includegraphics[width=12.5cm]{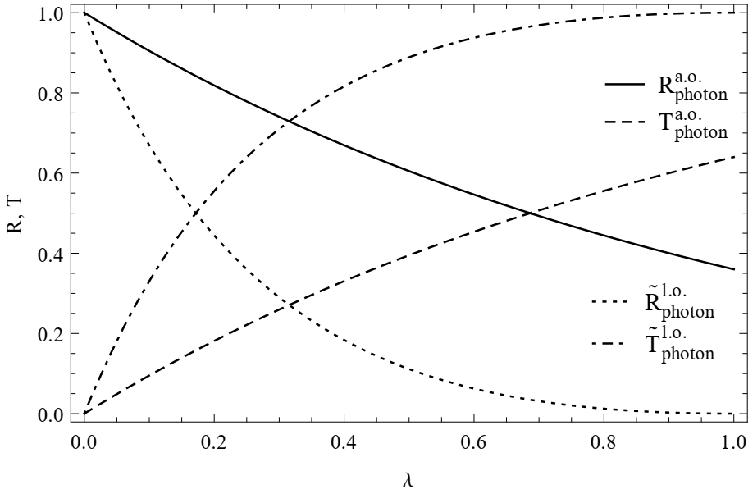}
\caption{Fresnel coefficients of DSWH for asymptotic and local observers.
\label{fig1}}
\end{figure}

At $\lambda =0$, the metric is just Schwarzschild. In this case, both sets of the Fresnel coefficients yield BH values: $R_{\text{photon}}^{\text{a.o.}}= \widetilde{R}_{\text{photon}}^{\text{l.o.}}=1$, $T_{\text{photon}}^{\text{a.o.}}=\widetilde{T}_{\text{photon}}^{\text{l.o.}}=0$, and the two observers identify the object as only a BH. These values represent a nice "effective optical medium" description of the usual curved spacetime geometric phenomenon, in which, to asymptotic observers, the photons would take infinite coordinate time to reach the horizon, thus appearing to them to stand still, i.e., \textit{not} \textit{transmitting} through the horizon\footnote{%
We are aware that, unlike the WH throat, the BH horizon is strictly not an ordinary surface, as it absorbs all incoming signals, but we alternatively describe it here as a barrier preventing transmission.}, which is equivalent to the statement $T=0$, and a BH is identified by both the observers. Second, any deviation from the set ($1,0$), which is introduced by $\lambda \neq 0$, however tiny, would mean that the observers would identify in varying degrees of probability a WH or a BH. In accordance with the arguments in Section \ref{sec:T_interpr}, the algorithm for identifying the probed object can be stated as follows:\textit{\ For }$0<\lambda <<1$, \textit{the observer measuring a relatively larger value of }$R$\textit{\ is more likely to identify the object as a BH. Similarly, the observer measuring relatively larger value of} $T$ \textit{is more likely to identify the object as a WH.} For instance, it follows from the above that $R_{\text{photon}}^{\text{a.o.}}>\widetilde{R}_{\text{photon}}^{\text{l.o.}}$ as shown in  Figure \ref{fig1}. Finally, though the Fresnel coefficients are valid for single probe, they are actually defined based on the \textit{averages} of many probes (Tangherlini's replica \cite{Tangherlini:1975}, with one photon for each replica). Therefore, it is not ruled out that the two observers might identify the object as the same (WH or BH) in some individual trials. Note, interestingly, that at $\lambda \rightarrow \infty $, the observation time $\Delta t\sim 2GM\ln \left( \frac{1}{\lambda ^{2}}\right) \rightarrow-\infty$, and with the tidal forces being finite (see Appendix \ref{appendixA}), Fresnel coefficients approach Schwarzschild values:  $R_{\text{photon}}^{\text{a.o.}}=\widetilde{R}_{\text{photon}}^{\text{l.o.}}\rightarrow 1$, $T_{\text{photon}}^{\text{a.o.}}=\widetilde{T}_{\text{photon}}^{\text{l.o.}}\rightarrow 0$.

%%%%%%%%%%%%%%%%%%%%%%%%%%%%%%%%%%%%%%%%%
\section{Conclusions}
\label{sec:concl}
%%%%%%%%%%%%%%%%%%%%%%%%%%%%%%%%%%%%%%%%%
This paper is deeply relevant to an important work by Lemos and Zaslavskii \cite{Lemos:2008}, where the authors argue that uncharged configurations such as DSWH (20) are bad mimickers of BHs, one of the reasons being the effect of infinitely large tidal effects on the moving body in the limit $\lambda\rightarrow 0$ (see Appendix \ref{appendixA}). This limiting divergence for DSWH, being in stark contrast to the non-divergent behavior of a BH horizon, allows the task of distinguishing black holes from their mimickers quite possible. The present paper clearly supports the conclusions in \cite{Lemos:2008}---it shows the possibility of this distinction, arguing from an entirely \textit{new} approach to probabilistic Fresnel coefficients. We showed that asymptotic observers are more likely to identify the probed object as a Schwarzschild BH, whereas near-throat local observers are more likely to identify it as a DSWH itself.

It should be emphasized that, to avoid divergent tidal effects experienced by a body on the throat, the original key assumption in \cite{Damour:2007} was \textit{not} to take the continuous limit $\lambda \rightarrow 0$ per se but to consider only \textit{discrete small values} of $\lambda $ to be used in the large coordinate time interval $\Delta t=2M\log \left(\frac{1}{\lambda ^{2}}\right) $ taken by light to connect the throat to the asymptotic region so that for a small enough $\lambda $, the two objects, the DSWH and the BH, would be indistinguishable within the large time-frame of observation. By the same token, for tiny \textit{non-zero} values, say in the order of $0<\lambda <<\exp \left(-10^{15}\right)$ \cite{Damour:2007}, the tidal forces could be very large but \textit{not infinite}. However, Lemos and Zaslavskii \cite{Lemos:2008} have usefully listed several categories of mimickers and found the extremal charged ones are the best mimickers, whereas gravastars are the worst, and the DSWH appears just above the gravastar category.

Thus, with the problem of divergent tidal effects out of the way for DSWHs, we studied the role of observers' locations by employing Tangherlini's approach to the probabilistic identification of the state of the system, whether a WH or a BH, in terms of Fresnel coefficients using photon probes by different observers---we do not ask what the object \textit{truly} is but ask how the object is likely to be \textit{observed} by differently located observers. The present methodology has the advantage that it can be extended to any static, spherically symmetric metric \cite{Nandi:2016}. The adaptation of a non-quantum indeterminacy \textit{\`{a} la} Tangherlini in the identification of the gravitating source by photon probes could have deeper implications for the semi-classical quantum gravity. \textit{Notwithstanding the semi-classical gravity, excluding BH solutions \cite{Barthiere:2018}, there is a non-zero probability that an observer may still identify the object as a BH, as argued above. This is the central result of this paper}.

However, we need to mention a caveat here: The effective medium approach is rather limited in scope, whereas Einstein's geometric approach is certainly far more versatile and in fact is a prerequisite for constructing the index itself from the metric. The medium approach could still be valuable in the sense that it allows one to borrow wisdom from the phenomena occurring in an ordinary optical medium---for instance, the Fizeau effect---and look for promising analogues in the "effective optical medium" representing a gravity field (see, for details, \cite{Nandi:2003}). An application of Fresnel coefficients in gravity was first used in detail in a different context in \cite{Nandi:2016}. The foundation for such application lay in the fact that light and particle kinematics in a real optical medium are remarkably analogous to those in the gravitational "effective optical medium." Then, Fermat's principle with refractive index $n(r)$, together with the "$F=ma$" formulation of Evans and Rosenquist \cite{Evans:1986, Rosenquist:1988, Evans:1990}, were previously shown to exactly reproduce the geodesic motion in the curved spacetime of general relativity \cite{Nandi:1995, Evans:2001, Evans:1996a, Evans:1996b, Alsing:2001, Alsing:1998}.

The main results from up to Section \ref{sec:app_to_DSWH} are embodied in Equations (28)--(31), from which it follows that the reflection and transmission coefficients of the DSWH throat are controlled solely by $\lambda$ at the DSWH throat ($r_{th}=M/2$). This means that, for $\lambda <<1$, asymptotic observers are more likely to identify the probed object as a Schwarzschild BH, since the reflection probability is larger, $R_{\text{photon}}^{\text{a.o.}}>\widetilde{R}_{\text{photon}}^{\text{l.o.}}$ (Equations (28) and (30)), but near-throat local observers are more likely to identify it as a WH, since the transmission probability is larger, $\widetilde{T}_{\text{photon}}^{\text{l.o.}}$ $>T_{\text{photon}}^{\text{a.o.}}$ (Equations (29) and (31)). Now, suppose that the asymptotic observer sends his/her information of the identification of a BH to the near-throat observer, who is identifying the same object as a WH. Then, the only way the latter observer can reconcile his/her observation of a WH with the received information of a BH (assuming it a dead or collapsed WH \cite{Gonzalez:2009, Bronnikov:2011, Bronnikov:2012, Bronnikov:2004, Bronnikov:2007}) is to conclude that the object he/she is identifying must be a \textit{ghost WH}. Once again, we remind ourselves that these identifications are based strictly on the probability for a very small but non-zero $\lambda$, as envisaged in \cite{Damour:2007}. No such ghost phenomenon appears in the case of an exact Schwarzschild BH for which $\lambda =0$, as shown in the text.

The support for the strategy in this paper from practical astrophysical observations, though it could be extremely interesting, is still lacking. However, we think that one has to look for exclusive WH signatures in an observed BH spectrum, such as in SgrA* or in M87. To this direction, accretion profiles and strong field lensing signatures are two good diagnostics for this purpose.

\appendix
\section[\appendixname~\thesection]{}
 \label{appendixA}
 
In the Lorentz-boosted orthonormal frame of freely falling observers, from the metric functions of DSWH (20), we get the curvature components as
\begin{eqnarray}
R_{\widehat{0}\widehat{1}\widehat{0}\widehat{1}} &=&-\frac{M\left\{2r^{2}\left( 1+\lambda ^{2}\right) -Mr\left( 8+5\lambda ^{2}\right) +8M^{2}\right\} }{r^{3}\left\{ r\left( 1+\lambda ^{2}\right) -2M\right\} ^{2}}, \\
R_{\widehat{0}\widehat{2}\widehat{0}\widehat{2}}^{(s)} &=&\frac{M\left(r-2M\right) }{r^{3}\left\{ r\left( 1+\lambda ^{2}\right) -2M\right\} }, \\
R_{\widehat{0}\widehat{2}\widehat{0}\widehat{2}}^{(ex)} &=&-\frac{M\lambda^{2}}{r^{2}\left\{ r\left( 1+\lambda ^{2}\right) -2M\right\} }\left( \frac{v^{2}}{1-v^{2}}\right) ,
\end{eqnarray}%
where $(s)$ denotes the value in the static orthonormal frame and $(ex)$ refers to excess curvature as measured in the moving orthonormal frame \cite{Horowitz:1997}. At the throat $r_{th}=2M$, these yield
\begin{eqnarray}
R_{\widehat{0}\widehat{1}\widehat{0}\widehat{1}} &=&-\frac{1}{16\lambda^{2}M^{2}},\quad \\
R_{\widehat{0}\widehat{2}\widehat{0}\widehat{2}}^{(s)} &=&0,\quad R_{\widehat{0}\widehat{2}\widehat{0}\widehat{2}}^{(ex)}=-\frac{1}{8M^{2}}\left( \frac{v^{2}}{1-v^{2}}\right) .
\end{eqnarray}%

It is quite clear only that $R_{\widehat{0}\widehat{1}\widehat{0}\widehat{1}} $ diverges at the throat in the limit $\lambda \rightarrow 0$, and so due to divergent tidal force
\begin{equation}
\Delta a_{\widehat{j}}=-R_{\widehat{0}\widehat{j}\widehat{0}\widehat{p}}\xi^{\widehat{p}},
\end{equation}%
a radially falling observer is infinitely elongated between toe to head, in contrast to the finite elongation of the horizon, but for a small value of $\lambda $, $R_{\widehat{0}\widehat{1}\widehat{0}\widehat{1}}$ is large but finite.

This feature persists even in the Bueno et al. \cite{Bueno:2018} form of the DSWH metric given by
\begin{equation}
ds^{2}=-\left( 1-\frac{2M}{r}\right) dt^{2}+\left( 1-\frac{2M\left(1+\lambda ^{2}\right) }{r}\right) ^{-1}dr^{2}+r^{2}\left( d\theta ^{2}+\sin^{2}{\theta }d\varphi ^{2}\right) .
\end{equation}%

This form also yields Schwarzschild BH at $\lambda =0$. The throat is now given by%
\begin{equation}
r_{th}=2M\left( 1+\lambda ^{2}\right) .
\end{equation}%

For this form, we find
\begin{eqnarray}
R_{\widehat{0}\widehat{1}\widehat{0}\widehat{1}} &=&-\frac{M\left\{2r^{2}-Mr\left( 8+5\lambda ^{2}\right) +8M^{2}\left( 1+\lambda ^{2}\right)
\right\} }{r^{3}\left( r-2M\right) ^{2}}, \\
R_{\widehat{0}\widehat{2}\widehat{0}\widehat{2}}^{(s)} &=&\frac{M\left\{r-2M\left( 1+\lambda ^{2}\right) \right\} }{r^{3}\left( r-2M\right) }, \\
R_{\widehat{0}\widehat{2}\widehat{0}\widehat{2}}^{(ex)} &=&-\frac{M\lambda^{2}}{r^{2}\left( r-2M\right) }\left( \frac{v^{2}}{1-v^{2}}\right) ,
\end{eqnarray}%
which, at the throat $r_{th}=2M\left( 1+\lambda ^{2}\right) $, yields
\begin{eqnarray}
R_{\widehat{0}\widehat{1}\widehat{0}\widehat{1}} &=&-\frac{1}{16M^{2}\lambda^{2}\left( 1+\lambda ^{2}\right) ^{2}}, \\
R_{\widehat{0}\widehat{2}\widehat{0}\widehat{2}}^{(s)} &=&0, \\
R_{\widehat{0}\widehat{2}\widehat{0}\widehat{2}}^{(ex)} &=&-\frac{1}{%
8M^{2}\left( 1+\lambda ^{2}\right) ^{2}}\left( \frac{v^{2}}{1-v^{2}}\right) .
\end{eqnarray}%

Once again, while the lateral components $R_{\widehat{0}\widehat{k}\widehat{0}\widehat{k}}$ of the Riemann curvature tensor together with their excesses are finite at the throat $r_{th}=2M\left( 1+\lambda ^{2}\right) ,$ the radial component $R_{\widehat{0}\widehat{1}\widehat{0}\widehat{1}}$ in the Lorentz-boosted frame can be arbitrarily large in the limit $\lambda\rightarrow 0$. Thus, the difference $\Delta a_{\widehat{j}}$ in radial tidal accelerations on two parts of a traveler's body would be infinitely large, and he/she would conclude that the throat is a singularity.

Note, however, that exactly at $\lambda =0$ in either metric (20) and (A7), one obtains the Schwarzschild BH, and we recover from the above expressions the well known components of the Riemann curvature tensor in the static or Lorentz-boosted frame:
\begin{eqnarray}
R_{\widehat{0}\widehat{1}\widehat{0}\widehat{1}} &=&-\frac{2M}{r^{3}}, \\
R_{\widehat{0}\widehat{2}\widehat{0}\widehat{2}}^{(s)} &=&\frac{M}{r^{3}}, \\
R_{\widehat{0}\widehat{2}\widehat{0}\widehat{2}}^{(ex)} &=&0.
\end{eqnarray}

\end{document}